# The mechanical design of SOXS for the NTT

Aliverti M.*[a], Hershko O.[b], Diner O.[b], Brucallassi A.[c], Pignata G.[d,e], Kuncarayakti H.[f,q], Bianco A.[a], Campana S.[a], Claudi R.[g], Schipani P.[h], Baruffolo A.[g], Ben-Ami S.[i], Biondi F.[g], Capasso G.[h], Cosentino R.[j,l], D'Alessio F.[k], D'Avanzo P.[a], Munari M.[l], Rubin A.[b], Scuderi S.[l], Vitali F.[k], Achrén J.[m], Araiza-Duran J.A.[d,e], Arcavi I.[n], Cappellaro E.[g], Colapietro M.[h], Della Valle M.[h], D'Orsi S.[h], Fantinel D.[g], Fynbo J.[o], Gal-Yam A.[b], Genoni M.[a], Hirvonen M.[p], Kotilainen J.[f], Kumar T.[q], Landoni M.[a], Lethi J.[p], Li Causi G.[r], Marafatto L.[g], Mattila S.[q], Pariani G.[a], Rappaport M.[b], Ricci D.[g], Riva M.[a], Salasnich B.[g], Smartt S.[s], Turatto M.[g], Zanmar Sanchez R.[l]

[a]INAF - Osservatorio Astronomico di Brera, Merate, Italy
[b]Weizmann Institute of Science, Rehovot, Israel
[c]ESO - European Southern Observatory, Garching, Germany
[d]Universidad Andres Bello, Santiago, Chile
[e]Millennium Institute of Astrophysics (MAS)
[f]FINCA - Finnish Centre for Astronomy with ESO, Turku, Finland
[g]INAF - Osservatorio Astronomico di Padova, Padua, Italy
[h]INAF - Osservatorio Astronomico di Capodimonte, Naples, Italy
[i]Harvard-Smithsonian Center for Astrophysics, Cambridge, USA
[j]INAF - Fundación Galileo Galilei, Breña Baja, Spain
[k]INAF - Osservatorio Astronomico di Roma, Rome, Italy
[l]INAF - Osservatorio Astrofisico di Catania, Catania, Italy
[m]Incident Angle Oy, Turku, Finland
[n]Tel Aviv University, Tel Aviv, Israel
[o]Dark Cosmology Centre, Copenhagen, Denmark
[p]ASRO - Aboa Space Research Oy, Turku, Finland
[q]University of Turku, Turku, Finland
[r]INAF - Istituto di Astrofisica e Planetologia Spaziali, Rome, Italy
[s]Queen's University Belfast, Belfast, UK

## ABSTRACT

SOXS (Son of X-shooter) is a wide band, medium resolution spectrograph for the ESO NTT with a first light expected in early 2021. The instrument will be composed by five semi-independent subsystems: a pre-slit Common Path (CP), an Acquisition Camera (AC), a Calibration Unit (CU), the NIR spectrograph, and the UV-VIS spectrograph. In this paper, we present the mechanical design of the subsystems, the kinematic mounts developed to simplify the final integration procedure and the maintenance. The concept of the CP and NIR optomechanical mounts developed for a simple pre-alignment procedure and for the thermal compensation of reflective and refractive elements will be shown.

## 1. INTRODUCTION

The SOXS instrument is a medium resolution spectrograph, focused on transients follow-up, to be installed at one of the NTT Nasmyth foci in La Silla in 2020.

*matteo.aliverti@inaf.it; phone +39 347 1504637; http://www.brera.inaf.it/

A detailed description of the instrument optical design is presented in [2], the control electronics in [3], the instrument control software in [4], and the assembly, integration, and test activities in [5].

In this paper the general mechanical design will be presented, starting from the corotator and the platform. All the optical subsystems (common path, calibration system, camera, UV-VIS spectrograph, and NIR spectrograph) will be installed on a flange and their mechanical design will be described in the subsequent chapters.

## 2. GENERAL LAYOUT

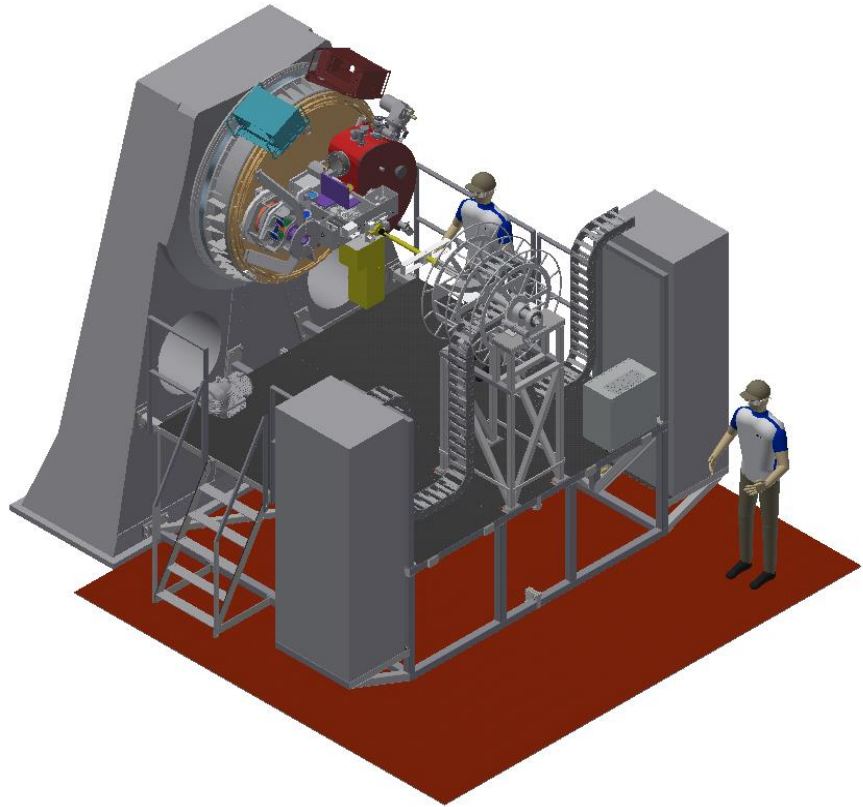

Figure 1: overall view of the SOXS spectrograph.

An overall view of the mechanical design of SOXS is shown in Figure 1. A **platform** is mounted using the connection points available on the NTT fork in order to keep it detached from the floor. To minimize the length of the cables going from the instrument to the electronic racks through the derotator, the racks have been directly mounted on the platform. Placing them on the floor would have required longer cables and a spring system to compensate for the phase difference of the telescope and the floor during azimuthal rotations.

A **corotator**, shown in Figure 2, is placed in the central part of the platform and is used to distribute all the cables coming from the instrument through the two chains or, in case of the liquid nitrogen line, using the central hole. On the right of Figure 2 the motor, the epicyclical reducer, and the spur gear used to move the corotator are shown together with the two bearings needed to support it. The feedback for the motor is provided by two linear potentiometers placed in the small box shown on the left. In the same box also a safety switch and a dragging system are installed. The first one is used to turn off both the instrument and the telescope derotator in case the motor is not able to keep the corotator aligned with the Nasmyth flange. Two springs are used to avoid any switch activation while the instrument is in stand-by mode. The dragging system is composed by two crown gears with a ±16° play that mechanically connect the corotator and the instrument in the unlikely event of a switch failure.

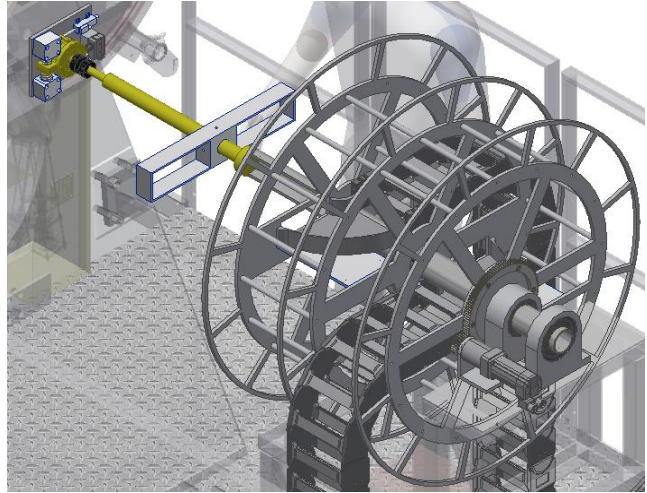
Figure 2: overall view of the corotator.

The instrument is mounted on the Nasmyth of the NTT as shown in Figure 3. An interface flange (orange) is used as a common support for the Common Path (CP, grey), the UV-VIS spectrograph (blue), and the NIR spectrograph (red).

The purple and the yellow elements are the camera and the calibration subsystems and are installed on the CP. To minimize the potential deformations related to the thermal dissipation of the New Generation Controllers (NGCs), they have been mounted outside the interface flange (light blue and dark red boxes).

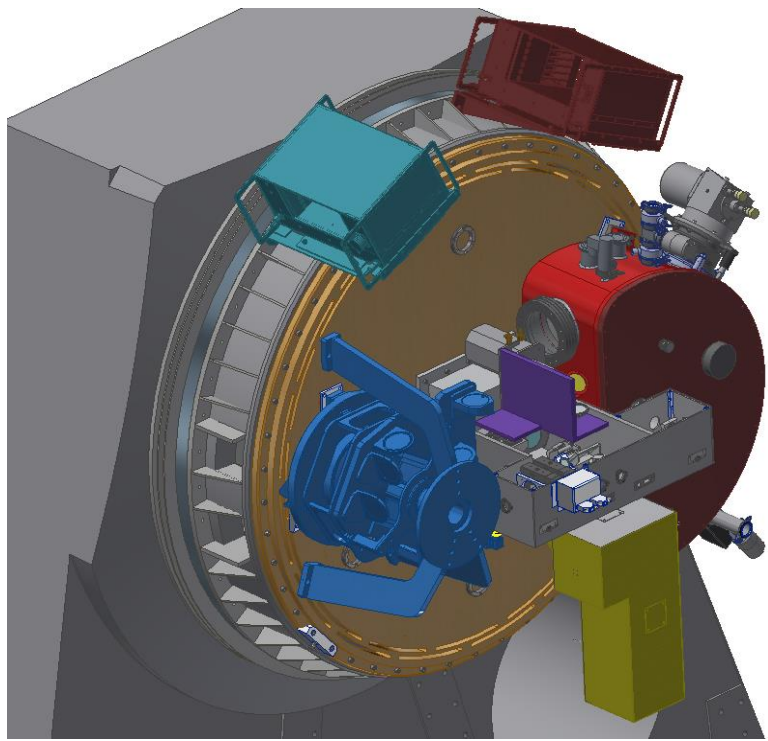
Figure 3: overall view of the instrument mounted on the Nasmyth flange.

## 2.1 Integration procedure and kinematic system

In order to decrease the integration time while mounting the different subsystems at the NTT, the two spectrographs and the CP are foreseen to be installed directly on the flange using a kinematic system. The flange is shown in Figure 4 with in light blue, grey, and dark red the support foreseen for the UV-VIS spectrograph, the CP and the NIR spectrograph, respectively.

All the elements, including the 55 mm thick flange, are made of Aluminium 6061-T6 in order to contain the weight and to have the same CTE of the optomechanical elements. As the NTT fork is made of steel, a radially symmetric flexure system has been foreseen on the flange. Those flexures allows for the compensation of the CTE difference removing the stresses and the membrane effect that would appear with a simple, rigid, flange without affecting the stiffness of the system.

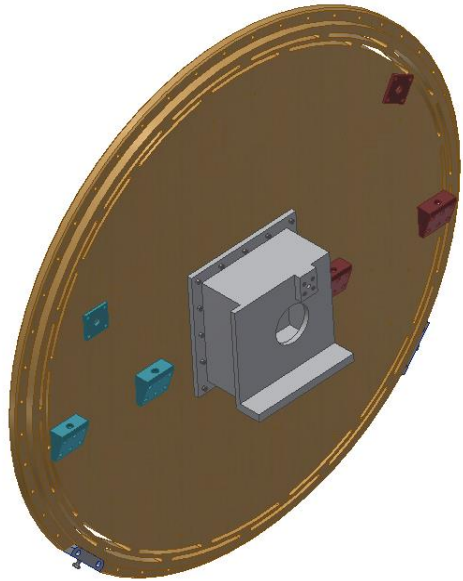

Figure 4: interface flange with the KMs supports (blue and red elements) and the regulation system (small plate along the perimeter of the flange).

The kinematic mounts (KMs) are shown in Figure 5 and have a design derived by the ones used in the ESPRESSO Front End [1]. Each of those elements is composed by 3 pieces made of high strength steel with a low friction coating applied on the contact surfaces: the bottom part is connected and mechanically aligned to the flange, while the top part is connected and mechanically aligned to the subsystem. A sphere, a cylinder, and a spherical washer are used to joint uniquely the two parts and a central screw is screwed through their centre in order to fix the subsystem and the flange together.

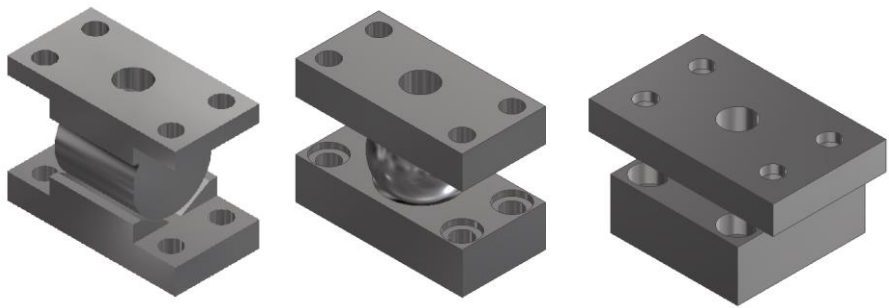

Figure 5: large KM system. From left to right: cylinder (2 dof), sphere (3 dof), and spherical washer (1 dof).

In Figure 6 an installation example shows the screws connecting the bottom part to the flange, the bolts connecting the upper part to the subsystem and the central element fixed with a screw.

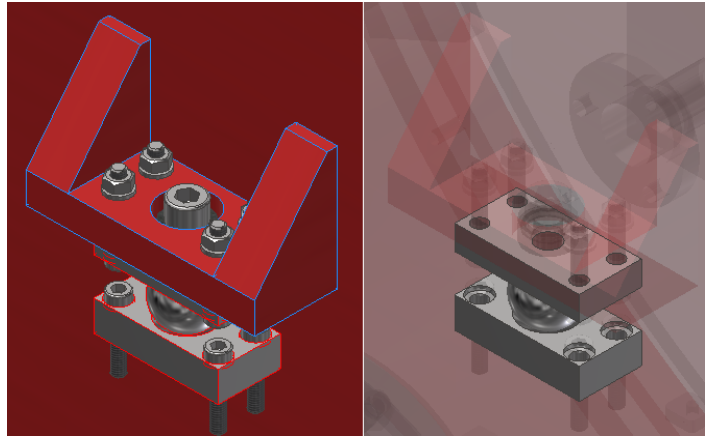

Figure 6: detail of one of the large KMs installed on the NIR spectrograph.

### 3. COMMON PATH

The purpose of the CP is to split the light coming from the telescope and inject it into the UV-VIS spectrograph and the NIR spectrograph, feed the AC with the light coming from the telescope or from the calibration box, and allow for the injection of the calibration light into the AC and/or the spectrographs. The SOXS CP is composed by a T shaped structure made of Aluminium 6061-T6 with an approximate size of 180x710x440 mm. The structure incudes one set of large KMs used as interface between CP and flange as presented in subsection 2.1.

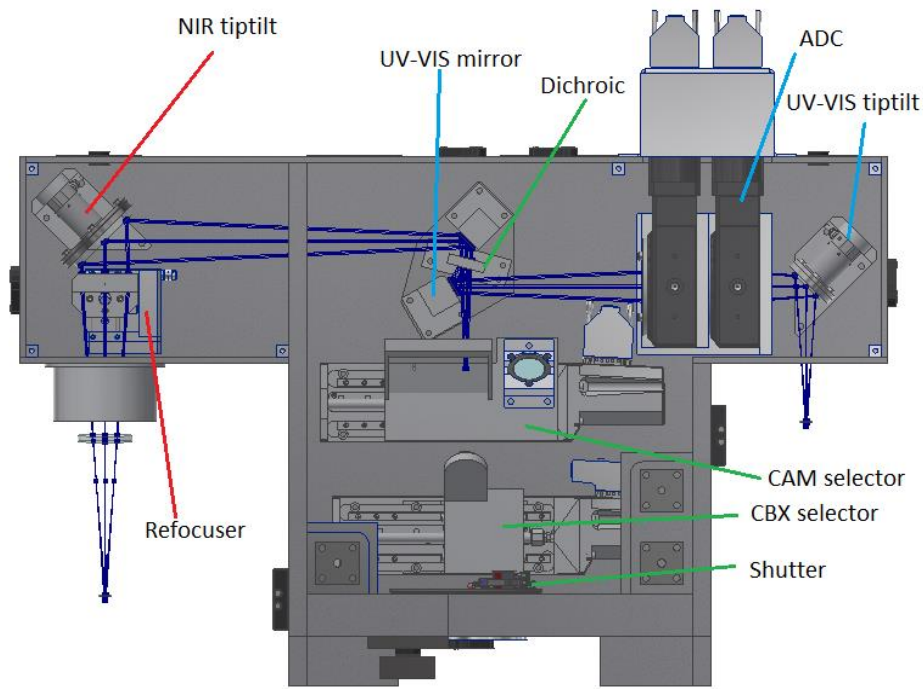

Figure 7: CP subsystem view. The optical beam enters from the bottom of the figure and is split in infrared (left) and visible (right) light.

The first part of the CP is composed by an instrument shutter, the calibration selector, the camera selector, and a dichroic. The light reflected from the dichroic will go to the UV-VIS spectrograph; in the CP, there is a flat mirror, an ADC with two counter-rotating prisms and a piezo tip-tilt. A small field lens is located above the slit and, even if from the optical point of view it is part of the CP, it is located inside the UV-VIS instrument and, therefore, will be presented from the mechanical point of view in the UV-VIS section.

The light transmitted by the dichroic goes to the NIR spectrograph; in the CP, there is a flat mirror, a piezo tip-tilt and a lens. A small field lens, the pupil stop and the NIR window are located above the slit and, even if from the optical point of view they are part of the CP, they are located inside the NIR instrument and, therefore, will be presented from the mechanical point of view in the NIR section.

The **shutter** is used both to shield the light and to protect the instrument when not in operation. It is a normally closed Vincent Associates Uniblitz CS65B (controlled with a Vincent Associates VED24 shutter driver) and it is screwed directly on the front surface of the CP. This shutter has a clear aperture of 65mm diameter while the beam is 55.8mm diameter.

The **CU selector** is composed by a flat mirror inserted or removed from the optical path using a PI DC motor stage with 102mm stroke (PI L-406.40DD10). Figure 7 shows the stage with the mirror positioning in calibration mode with the calibration light going into the instrument while. With the stage in position >53mm the telescope light is injected. This calibration stage shall be able to place the mirror with a maximum tilt of ±20 arcsec.

The **AC selection** uses the same linear stage to place a mirror with 2 holes in different positions (tilted +45°) or a pellicle (tilted -45°) on the focal plane. The selector provides an unvignetted FoV of 3.5' (38.1mm) to the AC. In addition to the previous tilt requirement of ±20 arcsec, this stage shall position the holes with a repeatability of 0.5 μm.

The **ADC** is made by two counter-rotating triplets. The optical elements will be purchased already mounted into barrel. The 2 barrels will be integrated and aligned into the 2 rotary stages (OWIS DMT100) separately. The 2 rotary stages will be relatively aligned over a single mount and, then, aligned into the main path.

The **tip-tilt mirrors** are used to correct small displacements of the spectrographs field and have a small support glued on the rotating part that allows for the mirror dismount in case of failure of the piezoelectric tip-tilt platform. The platform chosen is the PI S-330.2SL with 2mrad (~6.9 arcmin) stroke and 0.01arcsec resolution.

All the optical elements in the CP are mounted on the supports creating 3 3M 2216 glue spots on the back or along the radius of the element. This operation will be performed via small holes drilled inside the support.

A **refocuser**, moved with a PI M-111.1DG1linear stage, is also installed on the NIR side to compensate for relative defocuses between the UV-VIS and the NIR spectrographs.

In order to have a repeatable mount, 2 sets of small KMs are used as interface between the CP and the acquisition camera (AC) and the calibration (CU) subsystems.

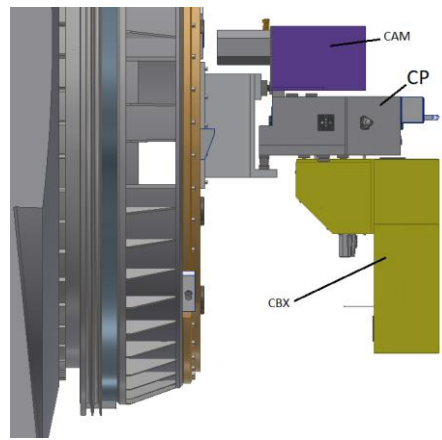

Figure 8: lateral view of the flange with its central column, the CP, the AC, and the CU subsystems. The KMs used to connect them are also shown in-between.

## 4. CALIBRATION UNIT

The calibration unit (CU) subsystem provides a synthetic light source for the purpose of wavelength and flux calibration of the observed spectrum. The light source is an integrating sphere equipped with quartz-tungsten-halogen (QTH), deuterium (D2), ThAr, and rare gas penray lamps. An optical relay system is used to direct the light from the integrating sphere to the telescope focal plane, emulating the F/11 beam of the telescope and providing uniform illumination across the spectrograph slit. The optical path is bent by two fold mirrors, with one of them acting as the calibration selector mirror inside the instrument CP. In addition to the calibration mode, the CU is capable to act as a star simulator, producing a point source using a pinhole mask unit positioned on an exchanger motor stage. This mask unit is equipped with additional optics to image the pinhole to the telescope focal plane (see [2]).

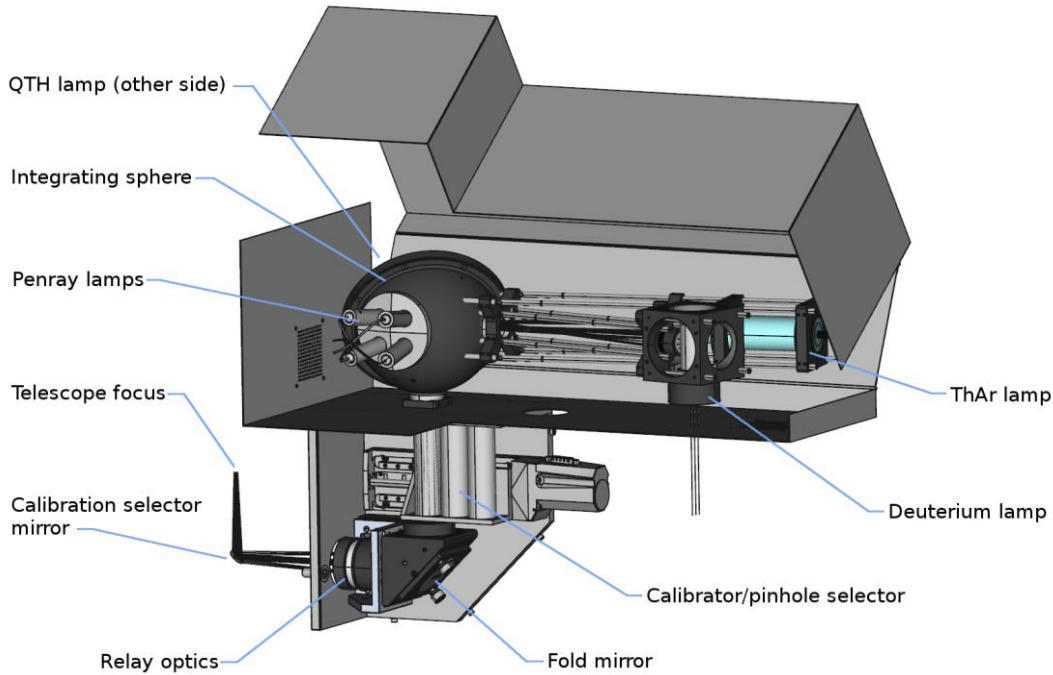

Figure 9: layout of the CU subsystem.

The parts of the CU include both commercial off-the-shelf and custom-made components. Integrating sphere from Labsphere, 4P-GPS-053-SL, is used. The sphere is Spectralon-coated, of 135 mm diameter, and has 4 ports. Three ports along the "equator" are used to hold the calibration lamps, and the fourth port at the "south pole" is used as the exit port. NeArHgXe penray lamps (Newport) are bundled together and occupy one port of the sphere. The QTH lamp (Osram) occupies another port across the penray port. The remaining port is used together by the ThAr (Photron) and D2 (Hamamatsu) lamps. The beams from the two lamps enter the sphere at a low angle, as greater angles mean smaller aperture at the sphere port. Support rods connect the ThAr and D2 lamps to the sphere, and they are wrapped with black aluminium foil around the optical path. The ThAr/D2 arm is about 450mm long. The CU optics and motor stage are attached to the inner wall of the optics enclosure box.

CU lamps are powered and controlled by the CU electronics that are placed separately in a rack. A safety interlock system is implemented on the cover of the lamp box. This interlock system will cut off electricity when the lamp box cover is opened, to ensure that no UV radiation from the lamps will harm the service personnel.

The whole subsystem will be enclosed in an aluminium case, with small drill holes grid near the position of the lamps for ventilation. These ventilation holes have diameters comparable to the thickness of the wall, making the beam angle of any scattered light steep enough, and are further covered by a plate on top of it to block leaked light. There is a long hinge at the upper back edge of the lamp box, and a supporting rail connects the back of the optics box and the lamp enclosure at the lower back edge. The dimensions of the lamp enclosure are 600 x 200 x 300 mm, and the optics box 200 x 200 x 100 mm. The CU is attached to the backside wall of the SOXS CP through a set of kinematic mounts. All cables pass through one hole on the underside of the lamp box wall.

## 5. ACQUISITION CAMERA

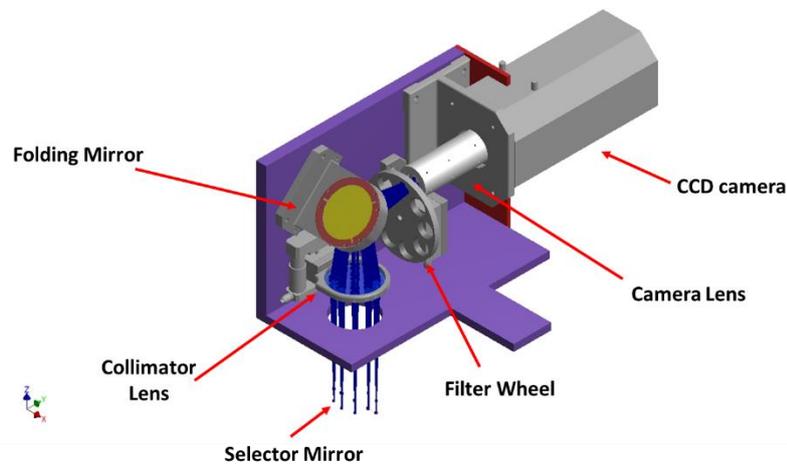

Figure 10: AC subsystem layout.

The AC is an essential tool with different functions [8]. It consists of a collimator lens, folding mirror, filter wheel, focal reducer optics, and CCD camera. The whole system is included in a structure made of 6061-T6 Aluminium.

The AC will receive an F/11 beam redirected from the telescope focal plane through the so-called AC selector. The latter is based on a sled that at the level of the Nasmyth focal plane, carries a single mirror with three positions for different functions and a pellicle. The mirror and the pellicle are tilted at 45° and direct light from sky or from the slits, respectively, to the AC optics.

The focal plane is placed at 500mm from the NTT Nasmyth interface and with a plate scale of 5.359 arcsec/mm. The selector will provide an unvignetted FoV of 3.5' to the AC.

The AC subsystem will be connected to the CP by a small KM system similar to the one presented in section 2.1.

The main functionalities of the AC system are described below, along with the corresponding AC selector positions:

- The CP flat 45° mirror shall select 3 positions:
  1. Acquisition and Imaging: the full field is reflected towards the AC.
  2. Monitoring: during the science exposure, the 45deg mirror is translated to place a hole on the optical axis. This passes an unvignetted field of 15arcsec to the spectrograph slits.
  3. Artificial star: the position is only used during daytime maintenance for alignment/centring.
- The CP slit viewing pellicle: the CP semi-transparent pellicle inclined towards the spectrograph allows using the AC system as a slit-viewing camera (with calibration lamp on).

### 5.1 Structure

The AC has a main T-shape structure made by aluminium 6061-T6 as shown in Figure 10. To ensure the maximum stiffness all the walls are structural with thicknesses of about 15mm. The weight of this part is about 1.4kg. The cover is not structural and is made of a 3mm thick Aluminium plate.

An Aluminium support will be mounted on the structure to fix the detector. This operation will be performed screwing it directly on the front surface of the structure. The centring of the CCD will be done shimming the 3 holes and 3 dowel pins used to place the CCD+support system.

## 5.2 Elements

Following the optical path, after the AC selector, a 60mm diameter lens (collimator) is foreseen to act as a re-focuser and will be placed on a linear stage with 15mm stroke (PI M-111.1DG1).

The lens is glued to an Aluminium mount by 3 3M 2216 glue spots.

The beam is redirected through a folding mirror (see Figure Figure 7 and Figure 10) along an axis parallel to the main telescope axis. The mirror will be mounted on a support screwed on the front surface of the external structure and kept in position inside the mount frame by axial leaf springs and a spring retainer.

In front of the camera lens there is a filter wheel made by a rotary stage (PI M-116.DG) and a custom filter support, which can host up to 9 elements.

The subsequent camera, formed by 2 doublets and 2 singlets of max diameter of 28 mm, relays the Nasmyth focus on the detector, with a F/#=3.6. The total length form focus to focus is 434.15 mm. For the camera lens, it is foreseen to include the 2 doublets and the 2 singlets in a tubular structure anchored to the camera mounting. The optical components will be mounted creating 3 3M 2216 glued spots on the side of the element. This operation will be performed via small holes radially drilled inside the support.

## 6. UV-VIS SPECTROGRAPH

A more complete description of the UV-VIS spectrograph can be found in [9]. From the mechanical point of view, the structure of the UV-VIS spectrograph must support the camera (three elements), the gratings, the dichroics, the OAP, the slit mechanism and slit-viewing camera, and the detector/cryostat assembly. The mechanical system is comprised of four parts which are connected via kinematic mounts: an Al6061 base which connects the UV-VIS system to the NTT flange, an Invar frame which holds camera primary mirror and slit mechanism, a stainless steel frame which holds the CaF2 corrector and field flattener, and an Al6061 plate which holds the feed small optics. The mechanical structure is shown in Figure 11. The subsystem is mounted on the flange with a set of KMs with the sphere and cylinder positioned along the long axis of the slit to ensure the slit location relative the NIR arm.

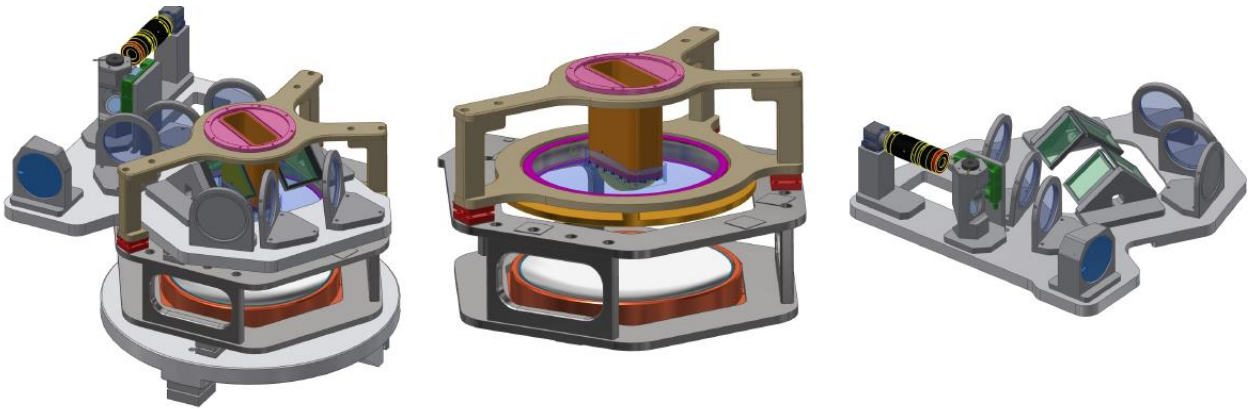

Figure 11: mechanical design of the UV-VIS spectrograph. From left to right: overall view, camera, and feed plate.

The camera mount composed by a stainless steel frame connected to the Invar frame with a three-cylinder KM. This mount ensures the camera centring, while the Invar frame ensures the distance between the corrector and the primary mirror is insensitive to thermal changes. The stainless steel frame also counteracts the extension of the stainless steel tube holding the field flattener. The feed plate holds the Al6061-T6 mounted dielectric, dichroic mirrors, and the off-axis parabola. A mechanism is mounted on the invar supports and is used for slit selection and to feed a slit-viewing camera through a prism.

# 7. NIR SPECTROGRAPH

The SOXS NIR spectrograph is composed by a D-shaped Vacuum Vessel (VV), a thermal shield and an optical bench. All those parts are made of 6061-T6 Aluminium. The VV includes one set of large KM used as interface between the NIR spectrograph and the flange. The SOXS NIR Spectrograph shall collect the light coming from the CP switching between different slit sizes (0.5", 1.0", 1.5", 5.0"). In addition to those slits, a pinhole (0.5") on the focal plane will be installed on the slit stage for alignment purposes.

The mass of the spectrograph is about 200 kg enclosed in a volume of about 900x700x400 mm (excluding the flanges and the cryo-vacuum elements connected to the flanges).

An overall view of the spectrograph is shown in Figure 12. On the right, the closed system shows the passive vacuum elements and the entrance windows. The VV weights around 85 kg and its base is the only structural part of the system and it is 25 mm thick to minimize bending between ambient and operating pressure while all the walls are 20 mm thick.

On the wall a total of 7 entrances are welded:

- 3 DN 160 CF flange to ease the access to the detector, to mount the TMP, and to install a window behind the grating for alignment purposes,
- 2 DN 50 ISO KF flanges to mount the feedthrough,
- 2 DN 25 ISO KF flanges to mount the pressure gauges,
- 1 DN 160 ISO K flange to mount the CCC system. Three small mounting turrets are also foreseen for the installation of the CCC damping system.

The CCC has a DN 160 ISO-K spring bellow keeping the vacuum between the VV ISO-K flange and another flange installed onto the CCC Leybold Coolpower 250 MD. This flange also provides the supports from three bars made of aluminium and three rubber dampers to decrease amplitude of the vibrations generated in the cold head. This concept is based on the one applied for TIMMI-2 and has the advantage of being compact and easily tuneable changing the rubber dampers and/or their preload.

Removing the cover and the upper part of the thermal shield all the optical elements are shown together with the cryopump.

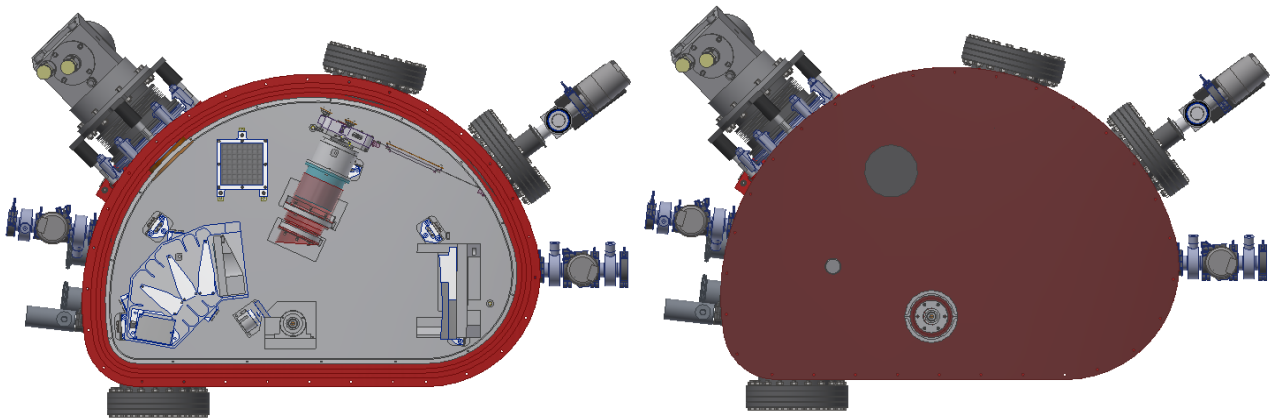

Figure 12: overall view of the NIR spectrograph. Left: internal view. Right: view with the VV closed.

The optical bench is supported by 6 flexures made of high strength steel. In order to ensure the stability of the slit location, a central pivot point is mounted right below the stage used to select the different available slits.

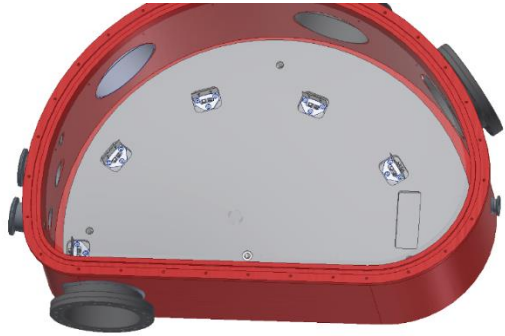

Figure 13: optical bench with the flexures and the central pivot.

**7.1 Optomechanical elements**

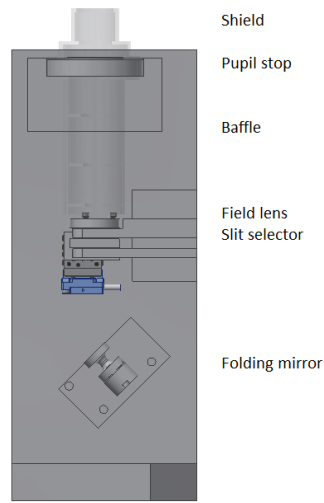

Figure 14: slit sub-bench.

The first elements encountered by the light coming from the CP are the pupil stop, the field lens, the slit selector, and the folding mirror. All those elements are mounted on an independent sub-bench pictured in Figure 14.

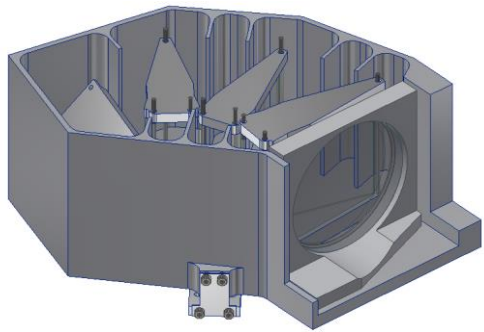

Figure 15: prisms sub-bench.

Another sub-bench, shown in Figure 15, is used to fix the relative position between the prism the grating, and the main lens, to improve their thermal stability, and to have a better baffling system.

One of the main peculiarities between the optomechanical mounts in the NIR spectrograph in terms of mechanical design is that no gluing is foreseen.

To do so, the mirrors are made of aluminium 6061-T6 and, to avoid residual stresses due to the mounting operation, and, will be manufactured with a wineglass foot shape (see Figure 16, left) or with integrated flexures. A similar concept has been already applied by ASTRON for the MIDI project [11].

All the lenses will be installed with a concept derived by the OMEGA2000 project [12]. The optomechanical are composed by two parts and the contact with the lens is done via a 45° chamfer on both sides of it. The lens is pressed between the two parts with three screws and three springs as shown in Figure 16 (right). In this way, the oblique forces provided by the springs will keep in position the lens during the warm up and cool down operations reducing the stresses related to the different CTE.

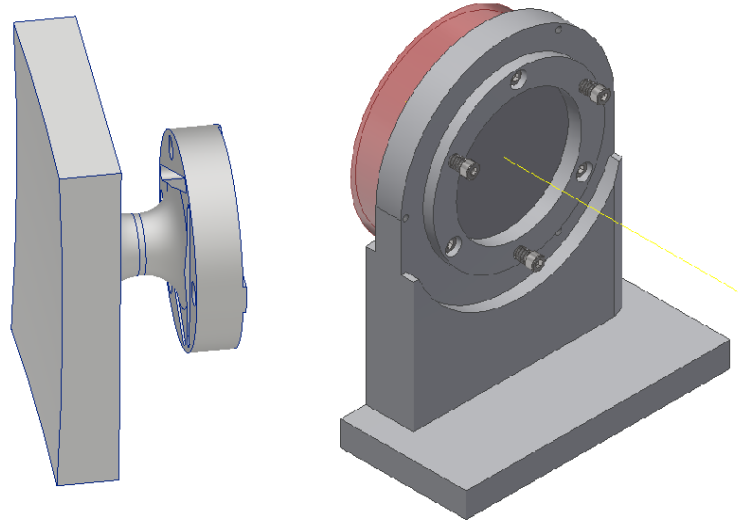

Figure 16: example of optical elements. Left: aluminium mirror. Right: lens support.

## 8. CONCLUSION

The SOXS project is going to be under Final Design Review in 2018. In order to avoid different behaviours of the subsystems and to keep the overall weight inside the mass budget, almost all the elements will be made of the same aluminium 6061-T6. Moreover, as the instrument is composed by five subsystems installed and tested in different locations, a modular concept based on kinematic supports has been developed. Following the final integration in Europe for the acceptance, the instrument will be installed at the NTT in La Silla with the first light expected by 2021.

## REFERENCES

[1] Schipani P, et al. SOXS: a wide band spectrograph to follow up the transients. Proc. SPIE 10702, (2018).
[2] Zanmar Sanchez R, et al. Optical design of the SOXS spectrograph for ESO NTT. Proc. SPIE 10702, (2018).
[3] Capasso G, et al. SOXS control electronics design. Proc. SPIE 10702, (2018).
[4] Ricci D, et al. Architecture of the SOXS instrument control software. Proc. SPIE 10702, (2018).
[5] Biondi F, et al. The Assembly Integration and Test activities for the new SOXS instrument at NTT. Proc. SPIE 10702, (2018).
[6] Aliverti M, Pariani G, Moschetti M, Riva M. Integration and alignment through mechanical measurements: the example of the ESPRESSO front-end units. In Ground-based and Airborne Instrumentation for Astronomy VI 2016 Aug 9 (Vol. 9908, p. 99087C). International Society for Optics and Photonics.
[7] Claudi R, et al. The common path of SOXS (son of x-shooter). Proc. SPIE 10702, (2018).
[8] Brucalassi A, et al. The acquisition camera system for SOXS at NTT. Proc. SPIE 10702, (2018).
[9] Rubin A, et al. MITS: the Multi-Imaging Transient Spectrograph for SOXS, Proc. SPIE 10702, (2018).

[10] Vitali F, et al. The NIR spectrograph for the new soxs instrument at the ESO-NTT. Proc. SPIE 10702, (2018).
[11] Kroes G, Kragt J, Navarro R, Elswijk E, Hanenburg H. Opto-mechanical design for transmission optics in cryogenic IR instrumentation. In Advanced Optical and Mechanical Technologies in Telescopes and Instrumentation 2008 Jul 23 (Vol. 7018, p. 70182D). International Society for Optics and Photonics.
[12] Baumeister H, Bizenberger P, Bayler-Jones CA, Kovács Z, Röser HJ, Rohloff RR. Cryogenic engineering for OMEGA2000: design and performance. InInstrument Design and Performance for Optical/Infrared Ground-based Telescopes 2003 Mar 7 (Vol. 4841, pp. 343-355). International Society for Optics and Photonics.